# QUANTIFYING MOVEMENT: EXPANDING THE ICHNOLOGIST TOOLKIT

Brittany A. Laing[1,2], Zoe Vestrum[3], Luke C. Strotz[4], M. Gabriela Mángano[1], Luis A. Buatois[1], Glenn A. Brock[2], and Lyndon Koens[5,6]

## 3.1 ABSTRACT

The trace-fossil record serves as a rich dataset to examine fossil behaviour, ecologic interactions at community level, and evolutionary trends in behaviour across geological time. Behavioural adaptations are often invoked in a variety of evolutionary hypotheses; however, few methods to quantitatively compare fossil behaviour exist. Movement paths, such as trails and trackways, are well-studied in extant-organism research where they are discretized and mathematically analyzed for behavioural strategies and trends. Here, we reference modern movement ecology research and present a methodology to discretize horizontal movement paths in the fossil record. We then demonstrate the utility of this methodology and the spatiotemporal data it collects via an analysis of the trilobite trace fossil *Cruziana semiplicata* and assess our results in light of three previous assertions about its recorded behaviour. Our analysis reveals the presence of three morphotypes, interpreted as three distinct behavioural variations, which persisted across multiple geographic localities and are interpreted to reflect changes in external conditions, internal states, or a combination of the two. Our research highlights the immense potential of this methodology to test behavioural hypotheses and provides an open-source groundwork for future research.

[1] *Department of Geological Sciences, University of Saskatchewan, Saskatoon, Saskatchewan, Canada.*
[2] *School of Natural Sciences, Macquarie University, Sydney, New South Wales, Australia.*
[3] *Department of Physics, University of Alberta, Edmonton, Alberta, Canada.*
[4] *State Key Laboratory of Continental Dynamics, Shaanxi Key Laboratory of Early Life & Environments and Department of Geology, Northwest University, Xi'an, China.*
[5] *Department of Mathematics, University of Hull, United Kingdom.*
[6] *School of Mathematical and Physical Sciences, Macquarie University, Sydney, New South Wales, Australia.*

## 3.2 INTRODUCTION

Ichnologic datasets are somewhat distinct from other fossil types as they provide glimpses of the interaction between an organism and its environment at a specific moment in time. The information trace fossils contain is thus truly multifaceted; they provide insights on an organism's internal behaviours and capabilities, how the surrounding environment impact those behaviours and capabilities, and how they may change through time. Trace fossils hence represent an important resource to test hypotheses which invoke any combination of these topics. Extricating information from trace-fossils is often not straightforward though, and current established methods rely on qualitative description and categorization. Trace fossils have been grouped and categorized on the main basis of behaviour (i.e. ethological classification), recurrent responses to environmental controls (i.e. ichnofacies), record of benthic communities (i.e. ichnocoenoses), life modes (i.e. ichnoguilds), and substrate reworking and tiering structures (i.e. ichnofabrics) (Seilacher, 1953; Frey & Pemberton, 1984; Bromley & Ekdale, 1986; Bromley, 1990; Taylor et al., 2003; Vallon et al., 2016; MacEachern & Bann, 2020). These qualitative tools have allowed for formidable strides in the field of ichnology and have deepened our understanding of ecologic interactions through Earth's history (Seilacher, 1967; Chamberlain, 1971; Ekdale & Bromley, 1983; Droser & Bottjer, 1993; Buatois et al., 1998). Yet there has been little work done to quantitatively analyze and compare behaviour with trace fossil data. Most quantitative work has focused on the simulation of fossil movement trajectories (Raup & Seilacher, 1969; Hofmann, 1990; Koy & Plotnick, 2010) or via reference to modern tracemakers (Miguez-Salas et al., 2022). Quantitative studies using trace fossil data are limited and either rely on the presence of easily identifiable morphologic features such as limbs and turns (Hofmann & Patel, 1989; Hofmann, 1990; Fan et al., 2017), focus on search strategies (Sims et al., 2014; Jensen et al., 2017), or investigate other features such as spatial distributions among populations (Pemberton & Frey, 1984; Mitchell et al., 2022).

This paper proposes a new method to address this gap. We take an actualistic approach, exporting concepts from movement ecology to mathematically analyze data-rich trace-fossil material relevant to movement paths. These fossil movement paths offer a prolific and relatively untapped data source for examining changes in behaviour through time (Sims et al., 2014). Movement paths are widely represented in the trace fossil record (e.g. trails and trackways), can be inferred via classic ichnological tools (e.g. ichnotaxobases, functional analysis), and possess quantifiable metrics that are well-studied in extant organisms. Similar trajectories and datasets are studied in an assortment of disciplines (e.g. ecology, biology, physics, physiology, data science, mathematics) within the paradigm of movement ecology and the potential application of the movement ecology framework to analyze trace fossil datasets has been previously discussed (Plotnick, 2012; Dorfman et al., 2023). Movement ecology research has experienced explosive growth in the last two to three decades and has recently seen the establishment of a dedicated scientific journal (Nathan & Giuggioli, 2013). The field encompasses movement paths made by vertebrates (including humans), invertebrates, plants, and microorganisms with recent trends in research covering topics such as dispersal, habitat selection, home ranges, foraging in marine megafauna, biomechanics, activity budgets, migration, breeding ecology, and human activity patterns (Joo et al., 2022).

Our method discretizes fossil movement trajectories (i.e. divides these curves into finite elements that can then be analyzed) and collects spatial and inferred temporal data along the path, aligning fossil organism movement data with extant organism movement data. This allows for the quantification, comparison, and analysis of fossil movement paths within or across specimens, ichnotaxa, facies, and geologic time. Here, we illustrate the applicability of the model to test previous assertions about the persistence and regional variation of the behaviour recorded by the trace fossil *Cruziana semiplicata*. We apply statistical techniques commonly used on similar extant organism movement data to test three behavioural hypotheses. Does temporary resting alter the future course taken by the *C. semiplicata* tracemaker? Do sets of paths from the same locality follow the same behaviour? Are there differences in the behaviour exhibited by *C. semiplicata* tracemakers in different geographic localities? We follow these questions up by examining the possibility of a sampling bias incurred by the locality-based grouping of *C. semiplicata* and broaden our investigation to examine the extent of variation in behaviour between individual specimens.

## 3.3 PREVIOUS WORK

Behavioural ecologists examine movement paths of organisms through the Movement Ecology Paradigm (Nathan et al., 2008). This paradigm offers a way to examine the process behind a movement path methodologically. It describes the formation of a movement path as a function of four factors: (1) The organism's intrinsic motivation to move (i.e. internal state), (2) the organism's ability to sense and respond to external signals (i.e. navigation capacity), (3) the organism's basic ability to move, both in terms of biomechanics and search strategies (i.e. motion capacity), and (4) The external factors which affect movement, such as nutrient distribution, predation, and competition (i.e. external factors). As the expression of these factors change, so will the resulting movement path. Simulations of movement paths can be conducted by dictating the expression of the four factors with explicit assumptions (e.g. Koy & Plotnick, 2007). Likewise, movement paths can be recorded and used to hypothesize on the expression of the four factors (Codling et al., 2008; Fan et al., 2017; Miguez-Salas et al., 2022).

Datasets for movement paths in extant organisms are composed of a series of spatial coordinates ($p = x, y,$ and $z$) associated with a time index ($t$). This series of ordered point data defines a series of vectors, or steps ($n$), which describes the movement path (Figure 3.1). The collection of this data can be coordinate-based or event-based (Long & Nelson, 2013). The former involves spatial coordinates collected at regular or irregular temporal intervals, regardless of how the organism was moving at the time, while the later defines the movement path as a series of spatiotemporal events delineated by the organism moving or remaining stationary (Spaccapietra et al., 2008). Movement data is often supplemented with data on the external environment, such as the location or shape of features that could affect the movement path (e.g. stream morphology, roads and paths, mountain ranges, food sources).

From this data, a variety of descriptive measures are commonly calculated including: step length ($d_n$), distance travelled in the X, Y, or Z directions ($\delta x_n$, $\delta y_n$, or $\delta z_n$), time lag ($t_{n+1} - t_n$), speed ($v$), absolute angle ($\alpha_n$), relative or turning angle ($\theta_n$), and displacement ($D_n$) (Fig. 1, Jones, 1977; Marsh & Jones, 1988; Calenge et al., 2009; Dray et al., 2010). From these measures, statistical analyses can be conducted to gain perspective on the expression of the four factors governing the formation of a movement path (ie. internal state, navigation capacity, motion capacity, and externa factors). Similar movement data collected from extant organisms have been used in analyses on orientation mechanisms, an organisms ability to track stimulants (Vickers & Backer, 1994; Svensson et al., 2014), the influence of landscape features (Boyce et al., 2010), presence of spatial and temporal memory (Dalziel et al., 2008), periodicity in movement behaviours (Boyce et al., 2010), spatial (i.e. home) range behaviour (Long & Nelson, 2013), and search strategies, among others (Calenge et al., 2009; Fagan et al., 2013; see Joo et al., 2022 for a review).

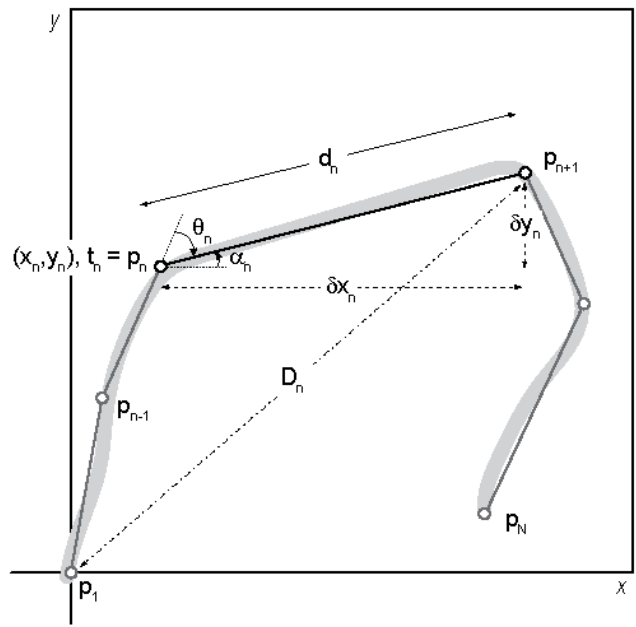

Figure 3.1. Sample movement path (thick grey line) with key measures annotated: p is the point data, with associated x- and y- coordinates (x, y) as well as inferred time (t) data; d is the step length; δx and δy are the distances travelled in the X and Y directions respectively; αn is the absolute angle measured relative to the X direction; θ is the turning angle; MSD is the mean squared displacement from the start point (p1) after n steps; D is the net displacement from the origin. Inspired by Marsh and Jones 1988 and Dray et al. 2010.

The probability distribution (i.e. PDF's) of turning angles provides a measure of the morphology of a movement path. These distributions can be used to detect the presence of movement patterns (Bartumeus et al., 2008; Long & Nelson, 2013). The means (or modes) of these distributions likewise reflect the morphology of the path, with means closer to 0 reflecting straighter trajectories (Cushman, 2010; Potdar et al., 2010; Figure 3.2A). More sinuous movement paths with an equal probability of left- and right- turns will yield bi- or tri- modally distributed turning angles which are symmetrical either side of 0 degrees (Figure 3.2B and 3.2C, Vickers & Baker, 1994). Asymmetry in turning angles distributions may indicate a preference in turning direction (Figcure 3.2D). The width of the distribution indicates how rounded turns are, with wider (or more spread out) distributions indicating more gradual changes in direction (Bartumeus et al., 2008).

59

Turning angles have been analyzed in horizontal trace fossil paths in the form of goniograms (Hofmann & Patel, 1989; Hofmann, 1990; Jensen, 2017). Goniograms plot the turning angle per unit length along a trace. They were proposed as a method for use in morphometric analysis, with periodicity in goniograms interpreted to reflect non-randomness in fossil trajectories (Hofmann & Patel, 1989; Hofmann, 1990).

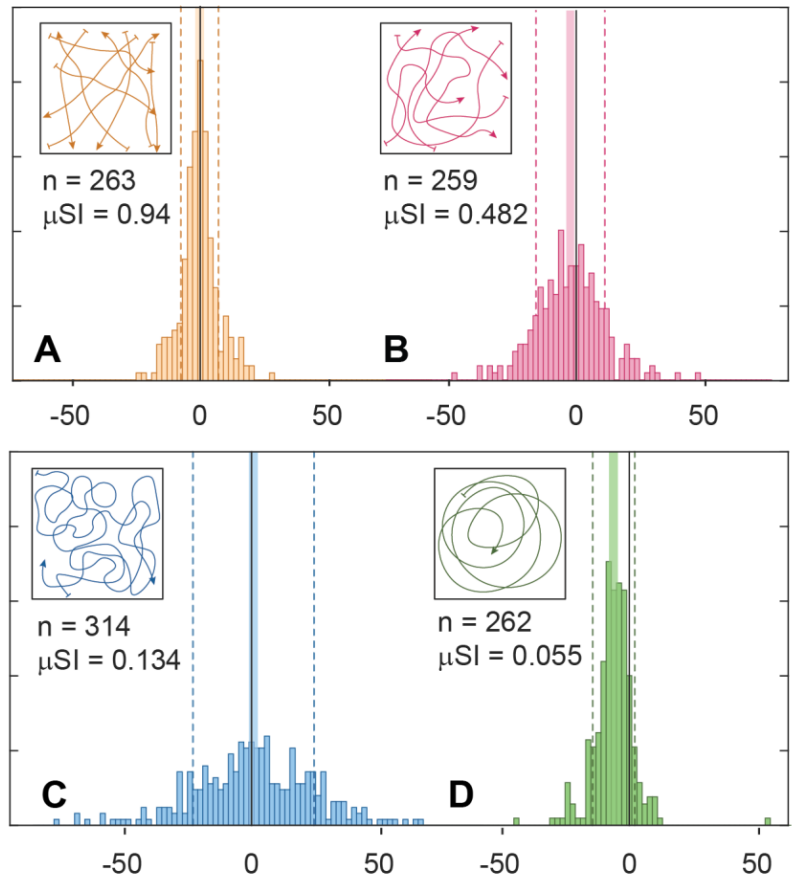

Figure 3.2. Sample movement path morphologies associated with turning angle probability distributions. **A**, a relatively straight movement path. **B**, a moderately sinuous movement path, **C**, a highly sinuous movement path. **D**, a movement path with a preferred turning direction. Thick vertical lines are the mean and dotted lines the standard deviation or turning angles. n indicates the number of angles collected per path morphology.

One common analysis performed on movement data is a two-sample *t*-test (i.e. Welch's *t*-test). This statistical test examines the likelihood that the population means of two samples were equal (Welch, 1947). These *t*-tests are often applied to determine if two samples differ significantly from each other and produce *p*-values to describe this difference. *P*-values reflect the probability of obtaining a sample mean difference at least as large as the observed difference, assuming the samples came from populations with equal means. Higher *p*-values, therefore, indicate a higher probability that the samples came from populations with the same means. *T*-tests and *p*-values are useful for hypothesis testing, with hypotheses that the two samples came from populations with equal ($H_0$, null hypothesis) or different ($H_a$, alternate hypothesis) means. The results of *t*-tests, the *p*-values, are discussed as a measure of the evidence against the null hypothesis ($H_0$), with smaller values indicating stronger evidence against $H_0$. One method to visualize the results of a series of two-sample *t*-tests is via a matrix of *p*-values. In these matrices, each cell contains the *p*-value result obtained from a single two-sample *t*-test. In this paper, these matrices will be shaded according to the strength of the evidence against $H_0$, with white indicating strong evidence ($p < 0.01$), light grey indicating moderate evidence ($0.05 > p > 0.01$), dark grey indicating weak evidence ($0.1 > p > 0.05$), and black indicating no evidence ($p > 0.1$).

## 3.4 CONSTRAINTS OF FOSSIL DATASETS & RATIONALE BEHIND PROPOSED METHODOLOGY

Two readily apparent constraints with the use of trace-fossil data to analyze organism movement are those of preservation and time (Plotnick, 2012). Movement is a spatiotemporal phenomenon and thus movement data requires the collection of both spatial and temporal data. In modern datasets, this is facilitated most often with wireless sensors, which transmits both the spatial and temporal component of a movement path. Trace fossils, however, only record organism-substrate interactions. Those movements which involve contact or manipulation of a substrate may be preserved, while those occurring solely in the water or air column will not. In addition, fossil datasets are notoriously time-averaged (Bromley & Ekdale, 1986; Savrda, 2016). This renders it difficult to infer the temporal domain associated with each movement path.

To resolve these issues, our methodology employs a theoretical velocity distribution of the movement path. By doing so, distances along the curvilinear length of the fossil path (i.e. the "segment distance" herein) can be theoretically related to a unit of time. While the precise speed of movement for a fossil tracemaker at any given time is unknown (though see Hsieh, 2020 and Hsieh et al., 2023 for speed estimates of modern analogues), the distribution of these speeds can be inferred. One solution is to apply an average velocity and assume that this is a reasonable approximation of the organism's velocity distribution. This allows for the fossil path to be subdivided into vectors (i.e. "segments") of uniform lengths which each represent the passage of a constant unit of time. This assumption is most reasonable if the organism is subjected to considerable constraints on their motion which would reduce the possible variability in their velocities. Trace fossils of infaunal or slow epifaunal foragers are good candidates for this assumption. These tracemakers move in near-constant contact with the substrate and the frictional forces imposed by this contact hinder their maneuverability and ability to rapidly modulate their velocity. While there will be some variability in the movement speeds of these organisms, the range of velocities is likely more constrained than in organisms that move freely on the sediment surface, in the water column, or in air, where there are considerably smaller frictional forces. Alternatively, the method proposed herein can accommodate fluctuation of movement speeds if evidence is available regarding the organism's movement velocity. The code for the segmentation of the discretized fossil path can simply be adjusted such that the segments created are non-uniform but rather follow another velocity distribution (i.e. faster movement equates to longer segments). Without further evidence about the movement speeds of the organism, however, the assumption of a reasonably average velocity is the simplest and most parsimonious solution and the one applied herein.

What, though, is a reasonable segment distance? Distances that have units (e.g. 3 mm or 1 cm) cannot be uniformly applied to all fossil paths as it does not account for changes in the size of the tracemaker nor does it relate to a measurement which the tracemaker can reference. The solution adopted in our methodology is to relate the segment distance to a readily available biologically based measurement—the width of the fossil path. For instance, rather than comparing data taken from coordinates spaced 1 mm apart for both *Climatichnites* Logan 1860 (average trail widths between 3 to 15 cm wide) and *Helminthoidichnites* Fitch 1850 (average trail widths between 0. to 3 mm wide), paths are compared relative to coordinates spaced *x* times the specimen's width apart (i.e. the "segment distance multiplier" in our methodology). An additional strength of this unitless approach is its resilience to fluctuations in trail width within populations as well as possible errors in scale. The segment distance multiplier can be set less than 1 (e.g. as 0.5, or half the trail width), though increasingly small segment distances incur similar issues with autocorrelation as increasingly small sampling intervals in extant organism data. As these intervals decrease, the dependence between the observed measures (e.g. step-length, turning angle) increase and introduce the issue of autocorrelation into the dataset (Dray et al., 2010). The segment distance multiplier is adaptable and can be set relative to any measure deemed relevant (e.g. jump length, organism length).

While three-dimensional movement within a substrate can be preserved in the fossil record, easily obtainable three-dimensional fossil paths are rare and difficult to collect in great numbers. As such, we suggest a focus on only those specimens that record bedding-parallel movement and are present on exposed slabs. In this manner, trace fossil paths can be reasonably recorded and interpreted in two-dimensions. The direction of travel, while sometimes able to be inferred with careful observation (e.g. via directional grooves created by appendages or muscular feet), is not often discernable in fossil movement paths. In these situations, path-independent measures, such as absolute position or mean displacement, will incur collection bias and can not be used in future analyses. However, these measures can be used (and will strengthen future analyses) if the direction of travel is discernable.

Our method explicitly focuses on directed horizontal movement that is preserved in the fossil record as trace fossils. This can be determined via the expression of ichnotaxobases and in turn the interpreted ethology. Criteria which would indicate a specimen is unsuitable for analysis with our method includes: the presence of a constructed wall or lining, zoned fill, true branching (i.e. secondary successive, primary successive, and simultaneous), and spreite (Buatois & Mángano, 2011). We suggest to restrict specimen selection to only those trace fossil paths that fulfill the following criteria: (1) the primary activity preserved in the specimen is locomotion or locomotion combined with feeding (i.e. repichnial or pascichnial trace fossil, respectively) (2) the specimen records bedding-parallel (2D) movement, and (3) the specimen was, or is able to be, photographed perpendicular to bedding with scale indicated.

## 3.5 METHODOLOGY

Our methodology involves collecting the spatial and temporal data associated with fossil movement paths. As proof of concept to demonstrate some utilities of this dataset, we focused on *Cruziana semiplicata* Salter 1853 from the late Cambrian (Furongian) to Early Ordovician (Tremadocian) (Supplementary Information, Figure 3.12). This ichnospecies was chosen as it is well-known in the field of ichnology and offered: (1) relatively contiguous movement paths, (2) a way to determine the direction of movement (i.e. v-shaped scratches), (3) multiple populations of roughly similar ages, and (4) previous tentative hypotheses which could be readily tested by our methodology. The specimens were grouped based on locality and include specimens of *C. semiplicata* from offshore environments of the Najerilla Formation of Spain (Álvaro et al., 2007; Seilacher, 2007, Pl. 14), subtidal environments of the Adam Formation of Oman (Fortey & Seilacher, 1997, Figure 1), shallow marine environments of the Olenus Beds of Poland (Radwański & Roniewicz, 1972, Figure 1), possible deep marine environments of the Kurchavinskaya Formation of Russia (Jensen et al., 2011, Figure 2 and Figure 3), and shallow marine environments of Wales (Crimes 1968; 1970, Plate 5). Perhaps the most well-known are the Spanish *C. semiplicata* from the Najerilla Formation (Figure 3.3), featured in Seilacher's Fossil Art exhibit (2008) and his seminal book Trace Fossil Analysis (2007, Plate 14). These specimens are found on two neighboring slabs and consist of four subgroups of movement paths, two moving counter-clockwise (i.e. Spain A & C) and two clockwise (i.e. Spain B & D). In his discussion, Seilacher (2007) postulated about the behaviour recorded by *C. semiplicata*, which we have formatted into hypotheses: (1) that the Spanish tracemakers followed a "fixed program" or stereotyped behaviour (*sensu* Wainwright et al., 2008) unaffected by temporary resting, (2) that all four Spanish subgroups were made by individuals with the same stereotyped behaviour, and (3) that further work may support separating *C. semiplicata* into two ichnosubspecies, representing two stereotyped behaviours. Our methodology offers the ability to test these hypotheses. As the morphology of movement paths are a product of organism behaviour, these datasets serve as a proxy to examine organism behaviour in the fossil record. Turning angle distributions quantitatively describe the morphology of movement paths and can be compared for statistical differences at the specimen-level (individual *C. semiplicata* trajectories), subgroup-level (sets of paths made by the same individual), and group-level (trajectories found in the same locality). We performed a series of two-sample *t*-tests to examine statistical differences in the means of turning angle distributions to test the hypotheses that the two-sample means (i.e. specimen, subgroup, or group turning angle distribution means) came from populations with equal ($H_0$, null hypothesis) or different ($H_a$, alternate hypothesis) means. If $H_0$ is supported, then we can infer that the two samples (i.e. specimens, subgroups, or groups) being compared demonstrate similar movement patterns. Conversely, if $H_0$ is rejected, then we can infer that the samples being compared demonstrate different movement patterns.

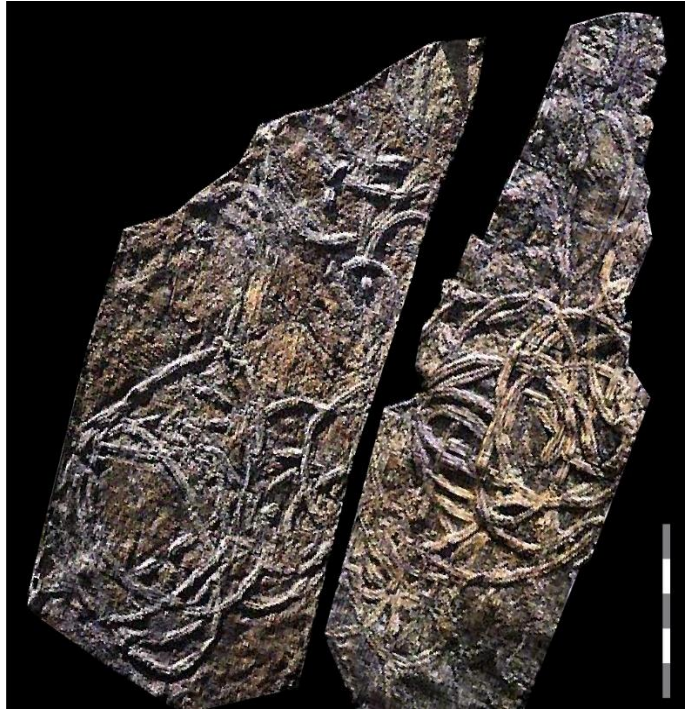

Figure 3.3. *Cruziana semiplicata* specimens from Spain. From Seilacher, 2008. Scale bar is 50 cm.

3.6 DATA COLLECTION

*3.6.1 Step A: Discretization*

The first step in the quantification of the fossil paths is their conversion to 2D curves that can be further analyzed in a numeric or statistical computing program (e.g. MATLAB or R). To this end, trace-fossil paths were first traced from photographs, digitized, and finally transformed into x- y- coordinates denoting a curve. Photographs of all specimens were brought into a vector graphics editor (Adobe Illustrator). Each fossil path was then traced using the pen tool, with the average width of the path preserved, and copied into its own artboard (Figure 3.4). Fossil paths were subdivided wherever strong evidence of a change of behaviour (i.e. movement phase, Nathan et al., 2008) was present, indicated by the presence of resting traces (e.g. *Rusophycus*). If travel direction was available, this was indicated via an arrow. For photographs containing multiple specimens, individual paths were indicated and numbered on a copy of the original photo for cross-referencing. Likewise, the artboards containing individual fossil paths were named according to the specimen number. Scale bars were also traced and indicated on every specimen's artboard. These individual specimen path images were then exported as an image file, ready to be imported into the computing program.

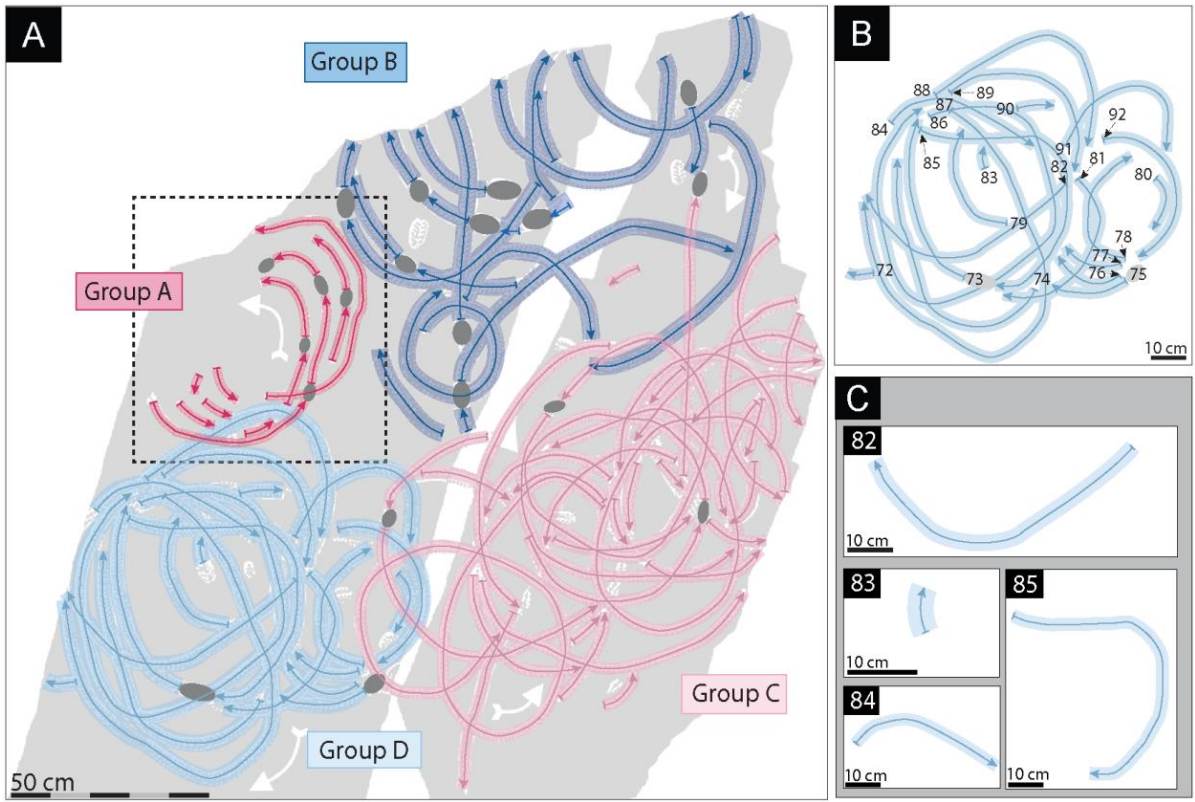

Figure 3.4. Sample set up for the creation of individual movement paths from a single fossil slab. A, the entire fossil slab with all specimens of *Cruziana semiplicata* denoted by lines (*Rusophycus* specimens are denoted by solid grey ovals), with colours used to indicate different population subgroups. Original drawing of trace-fossil slab by Seilacher (2007). Inset is shown in B. B, all trails of the light blue subgroup, numbered with their specimen ID. C, individual specimens, numbered, with scale indicated. These are saved as individual image files, ready to be imported into MATLAB.

Each specimen path image was then imported into MATLAB and loaded into a coordinate system dictated by the pixel size of the image. Points within this coordinate system were then collected to denote the start and end of both the scale bar and average width of the path, and their x- and y- coordinates saved into MATLAB arrays. Similarly, points along the fossil path were selected sequentially, and their x- and y- coordinates saved in another MATLAB array. If travel direction was discernable, the points were selected from start to end of the fossil path. To verify that the points selected provide an adequate representation of the fossil path, scale, and width, they were plotted and superimposed on the specimen path image (Supplementary Information, Figure 3.11). If the points were erroneously located outside of the path, the process would either be repeated or, if only one or two points were outliers, these coordinate(s) were removed from the MATLAB array. As the real distance between the two points on the scale bar are known, this enabled a scale ratio (Supplementary Information, Table 3.3) to be calculated and subsequently used to convert the pixel-coordinate system into one that reflects real-world measurements. At this point, an approximation of the curvilinear distance along the fossil path was calculated and saved for each specimen (Supplementary Information Table 3.3). This was done via the summation of all Pythagorean distances between neighbouring points along the fossil path.

### 3.6.2 Step B: Segmentation

At this stage, the points along the fossil path are indiscriminately spaced and used only to define the spatial domain of the movement path. The collection of coordinate-based movement data, with both an associated spatial and temporal domain, is possible if we can know, or can reasonably assume, the velocity distribution of the organism's movement (see discussion in "Constraints of fossil datasets"). In the case of *Cruziana semiplicata,* the organism moved with near-constant contact with the substrate and demonstrated little change in its mechanics of motion and so we applied an average velocity as a reasonable approximation the velocity distribution. As a result, a set distance $d$ (i.e. "segment distance" herein) along the curvilinear length is equivalent to a unit time. Our method calculates the segment distance as a function of the trail width collected in Step A, both to ground the segmentation in a biologic feature and to avoid any potential issues with scale in the original photographs (see discussion in "Constraints of fossil datasets"). Once a segment multiplier was chosen and uploaded into MATLAB, coordinates spaced at the resulting segment distance apart along the fossil path curve were interpolated. The MATLAB arrays of these coordinates were saved, with the chosen segment distance indicated, for future analysis. These arrays form the basis for subsequent analyses (e.g turning angle PDFs, *t*-tests, SI calculations, etc.).

### 3.6.3 Step C: Turning angle

The turning angle ($\theta$) at $t_0$ is defined as the angle between $t_{-1}$, $t_0$ and $t_{+1}$ and represents the deviation in travel direction from the preceding step (Figure 1). Mathematically, each point of interest ($t_{-1}$, $t_0$ and $t_{+1}$) can be defined as vectors relative to the origin ($\overrightarrow{t_{-1}}$, $\overrightarrow{t_0}$, and $\overrightarrow{t_{+1}}$). This allows for the vectors connecting points $t_{-1}$, to $t_0$ ($\overrightarrow{n_1}$) and points $t_0$ to $t_{+1}$ ($\overrightarrow{n_2}$) to be determined through simple vector subtraction,

$$\vec{n_1} = \vec{t_0} - \vec{t_{-1}},$$

$$\vec{n_2} = \vec{t_{+1}} - \vec{t_0}.$$

The corresponding unit vectors were then calculated by dividing these movement vectors by their Euclidean norms, such that such that $\hat{n}_1$ is a vector with the same direction as $\vec{n_1}$, but whose magnitude is equal to 1,

$$\hat{n}_1 = \frac{\vec{n_1}}{||\vec{n_1}||}.$$

The angle between these unit vectors can then be readily determined by calculating the arctan2 of the cross product of the normalized vectors divided by the dot product of the normalized vectors,

$$\theta = arctan2(\hat{n}_1 \times \hat{n}_2, \hat{n}_1 \cdot \hat{n}_2).$$

Data arrays containing these angles, and the temporal domain associated with them, were created and saved for each specimen and at each segment distance multiplier.

3.7 HYPOTHESES

*3.7.1 Does temporary resting alter the future course taken?*

Interspersed between the trails of *C. semiplicata* on the Spain slabs are deep rusophyciform depressions. These indicate a transition between discrete movement phases, from horizontal movement to shallow downwards burrowing, and are interpreted as resting (i.e. cubichnial) traces. The formation of these depressions has been suggested to have no impact on the trajectory of subsequent *C. semiplicata* paths and has been interpreted by the author to reveal that the tracemaker followed the same stereotyped behaviour over a considerable amount of time (Seilacher, 2007). If the specimens are maintaining a similar movement pattern after forming these depressions, then two-sample *t*-tests of the specimens should reveal no evidence to reject H₀. To examine this hypothesis, we ran specimen vs. specimen two-sample *t*-tests (i.e. Welch's *t*-test) for each pair of rusophyciform-divided Spain specimens (Table 3.1).

*3.7.2 Do sets of paths from the same locality follow the same behaviour?*

The Spanish specimens of *C. semiplicata* are conveniently located on two adjoining slabs and contain four sets of trails of differing widths (Figure 3.3 & 3.4). While two sets turn predominately left and two predominately right, all four exhibit a tendency to create large steadily curving trails which often over cross. This has been suggested by Seilacher (2007, p. 40) to represent a potential "circular scribbling" behaviour, inferred as a stereotyped behaviour to forage nutrient-rich regions more efficiently, employed by all four tracemakers. To examine the difference between the turning angle datasets of the four sets of paths (i.e. subgroups) we created PDFs of the turning angles for each subgroup and calculated the means, variances, and number of angles collected. Next, we preformed two series of subgroup vs. subgroup two-sample *t*-tests (i.e. Welch's *t*-test). If all four subgroups record the same movement pattern, then the *t*-tests should reveal no evidence to reject $H_0$ ($p > 0.1$) between all four Spanish subgroups.

*3.7.3 Are there differences in behaviour by geographic localities?*

It was suggested by Seilacher (2007) that *Cruziana semiplicata* might be subdivided into ichnosubspecies based on linear and scribbling behaviours (Seilacher, 2007). Specimens from Oman were likened to those from Spain and attributed to the "circular scribbling" behaviour, while specimens from Wales were described as a straighter group (Seilacher 2007, p. 40). To examine if *C. semiplicata* does demonstrate different behaviours in different localities (i.e. groups herein), we examined their grouped turning angle distributions. Like our analysis of the Spanish subgroups, we first examined turning angle for each group and calculated their means and variances. To examine differences between the turning-angle distributions, we conducted 2 series of group vs. group two-sample *t*-tests (i.e. Welch's *t*-test). If there are differences in the movement patterns recorded by the Spanish and Omani groups, then these *t*-tests should reveal: (1) no evidence to reject $H_0$ ($p > 0.1$) between the Spain and Oman groups and (2) evidence to reject $H_0$ ($p < 0.1$) between the Wales group and the Spain and Oman groups.

*3.7.4 How does sampling bias affect these results?*

Previous inferences and hypotheses on the behavioural patterns demonstrated by *C. semiplicata*, and our aforementioned tests of these hypotheses, are discussed in terms of locality-based groups. This grouping criteria could impose a sampling bias on our *t*-test results. To investigate if the results obtained from our analyses represented a true behavioural signal, we looked for similarities and differences between individual specimens of *C. semiplicata*. This was achieved through a series of specimen-vs-specimen two-sample *t*-tests between all *C. semiplicata* specimens. We developed a threshold argument to isolate specimens whose movement patterns were distinguishable from the majority of *C. semiplicata* specimens (i.e. "different" specimens). Our argument included (1) a threshold *p*-value which we deemed to be sufficient enough to reject $H_0$ (i.e. significantly "different" enough) and (2) a threshold percentage of specimens from which an individual specimen was significantly different from. To determine if these specimens were having a significant impact on the results of our group-vs-group two-sample *t*-tests, we re-ran these series of *t*-tests, this time excluding the turning angle data from the "different" specimens (Fig . 5C and 7C). We chose a *p*-value less than 0.1 as a sufficient probability to reject $H_0$ between specimens. For each specimen, we then summed how many *p*-values were greater than our threshold *p*-value (i.e. $p > 0.1$) and sorted our matrix by these summed amounts.

3.8 RESULTS

*3.8.1 Does temporary resting alter the future course taken?*

Specimen-vs-specimen two-sample *t*-tests revealed only two pairs of rusophyciform-divided Spanish specimens with evidence to reject $H_0$. Nine pairs showed no evidence to reject $H_0$ and 3 pairs did not possess enough turning angle data (i.e. were too short) to get a *p*-value. These results suggest on average a similarity between specimens before and after the rusophyciform depressions.

*3.8.2 Do sets of paths from the same locality follow the same behaviour?*

Analysis of the Spanish subgroup turning angle data revealed left-angle means for the left-leaning subgroups (Spain A & C) and vice-versa for the right-leaning subgroups (Spain B & D). All subgroups had means under 5 degrees and variances under 10 degrees, with right-leaning subgroups having slightly higher means and variances than their left-leaning counterparts.

Table 3.1. Results of specimen-vs-specimen two-sample *t*-tests of rusophyciform-divided Spain specimens, performed on direction-adjusted turning angle data (seg. mult. = 0.5).

| Specimen before | Specimen after | *p*-value |
|---|---|---|
| Spain 1 | Spain 9 | 0.2315 |
| Spain 9 | Spain 13 | 0.0395** |
| Spain 8 | Spain 10 | 0.0314** |
| Spain 11 | Spain 12 | 0.1199 |
| Spain 15 | Spain 16 | NaN |
| Spain 20 | Spain 21 | 0.8667 |
| Spain 23 | Spain 24 | NaN |
| Spain 24 | Spain 25 | NaN |
| Spain 25 | Spain 26 | 0.8705 |
| Spain 28 | Spain 22 | 0.8622 |
| Spain 43 | Spain 42 | 0.2687 |
| Spain 45 | Spain 56 | 0.9071 |
| Spain 56 | Spain 63 | 0.4996 |
| Spain 85 | Spain 73 | 0.4722 |

The first series of *t*-tests were performed on the unadjusted subgroup turning angle data (Figure 3.5A). This analysis found no evidence to reject $H_0$ for Subgroups A and C ($p = 0.5715$) nor for Subgroups B and D ($p = 0.1863$). There was, however, strong evidence to reject $H_0$ between the left- and right- turning subgroups ($p = 3.7 \times 10^{-37}$, $36.2 \times 10^{-62}$, $4.8 \times 10^{-43}$ and $7.3 \times 10^{-73}$). This result is expected, considering the strong directional bias in the turning angle of these subgroups. However, left- and right- turning subgroups could be coming from populations with identical absolute means, just with different preferred turning directions (i.e. $\mu = -3$ and $3$), possibly an artefact of behavioural asymmetry (Babcock, 1993). To address this, we corrected the turning-angle data for preferred turning-direction. The phi data for each specimen was multiplied by either -1 or 1 such that all specimen means were negative values. Subgroup vs. subgroup two sample *t*-tests were then redone with this direction-adjusted data and produced similar, though less extreme, results (Figure 3.5B). This suggests a similarity within left-turning subgroups and right-turning subgroups, but a distinction between the two even after accounting for trail width (see Methodology) and preferred turning direction. Consequently, the subgroups were grouped for future location-based analysis into groups Spain L, comprised of subgroups Spain A and C, and Spain R, comprised of subgroups Spain B and D, respectively (Supplementary Information, Table 3.2).

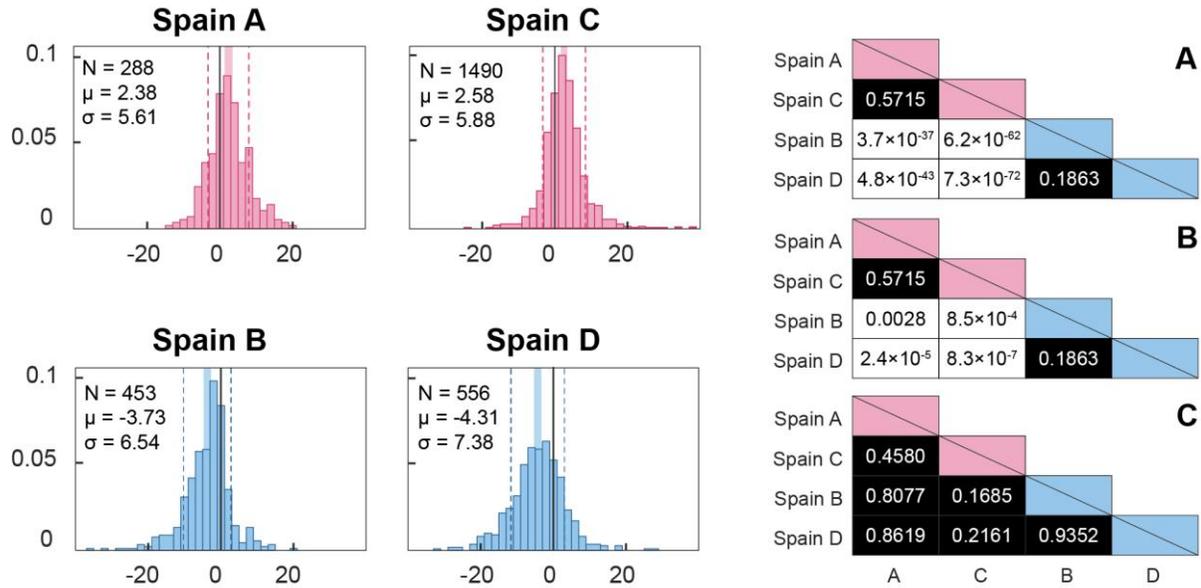

Figure 3.5. Top row: Turning angle (θ) PDFs of all subgroups of C. semiplicata from Spain (segment multiplier = 0.5). N = number of samples, μ = sample mean (thick transparent line), σ = sample standard deviation (thin dotted lines). Bottom row: Results of subgroup-vs-subgroup two-sample t-tests for all subgroups of C. semiplicata from Spain, performed on:  A: subgroup turning angle data, not adjusted for preferential turning direction. B: subgroup turning angle data, adjusted for preferential turning direction. C: subgroup turning angle data, adjusted for preferential turning direction with morphological outliers removed. Shade of boxes reflect p-value ranges: black = no evidence (p > 0.1) and white = strong evidence (p < 0.01).

*3.8.3 Are there differences in behaviourby geographic localities?*

Like our analysis of the subgroups of *C. semiplicata* from Spain, we first examined the turning angle PDFs for each group and calculated their means and variances (Figure 3.6). The means for Spain L, Oman, and Russia were left angles while Spain R, Poland, and Wales had right angle means. Notably, Russia had a much higher variance ($\sigma = 15.20$) than the other groups, which all had variances between 5.84 and 8.75.

The first series of *t*-tests compared the unadjusted grouped turning-angle datasets (Figure 3.7A). There was strong evidence to reject $H_0$ between most left-leaning groups (i.e. Spain L, Oman, and Russia) and most right-leaning groups (i.e. Spain R, Poland, Wales). Once again, we corrected the turning-angle datasets for preferred turning-direction and reran the group vs. group *t*-tests (Figure 3.7B). This subsequent analysis showed strong evidence to reject $H_0$ between the Spain R group and the Spain L, Oman, and Wales groups. There was weak evidence to reject $H_0$ for the Wales and Spain R groups and no evidence to reject $H_0$ between the Oman and Spain L groups.

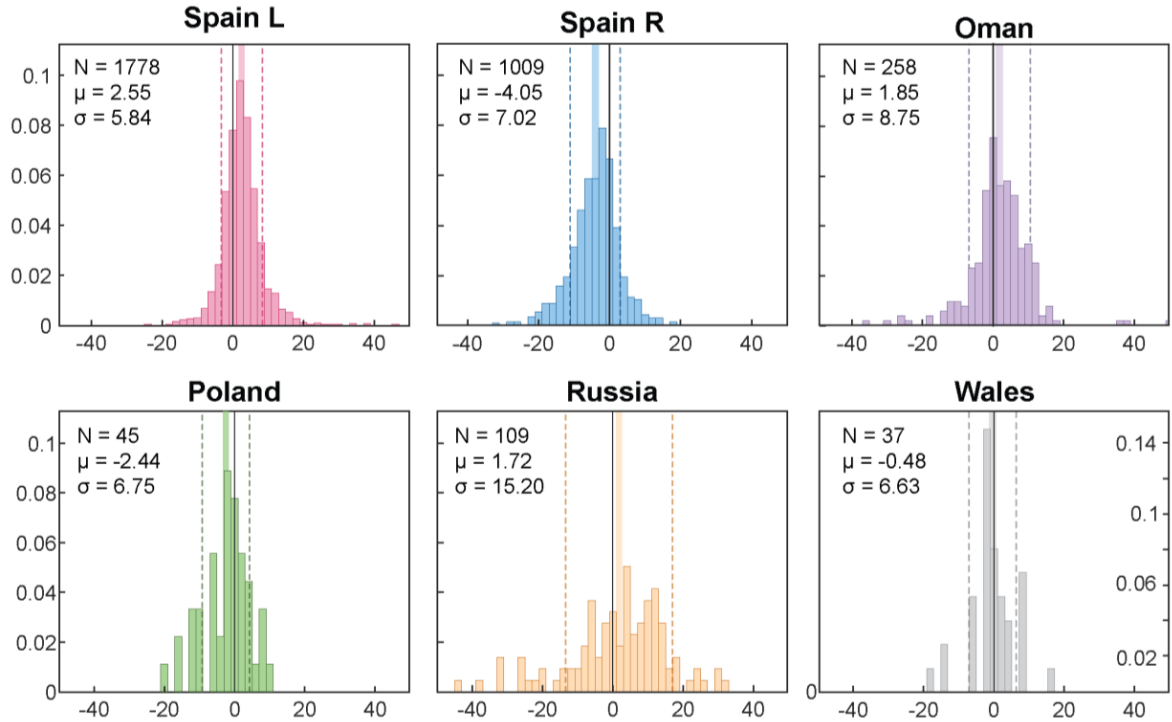

Figure 3.6. Turning angle (θ) PDFs for each studied group. N = number of samples, μ = sample mean (thick transparent line), σ = sample standard deviation (thin dotted lines).

*3.8.4 How does sampling bias affect these results?*

The results from our specimen-vs-specimen two-sample *t*-test revealed that topmost specimens were largely similar (i.e. *p* > 0.1) to each other (Supplementary Information, Figure 3.13). Additionally, we found that a threshold percentage of 25% adequately encompassed the remaining specimens. This resulted in 40 "different" specimens. To determine if these specimens were having a significant impact on the results of our subgroup-vs-subgroup and group-vs-group two-sample *t*-tests we re-ran these series of *t*-tests, this time excluding the turning angle data from the "different" specimens (Figure 3.5C & 3.7C). This analysis showed there was no longer evidence to reject $H_0$ between the four Spanish subgroups (*p* > 0.1, Figure 3.5C), nor between the Spain L and R groups (*p* = 0.1074, Figure 3.7C), nor for the Wales and Spain L groups (*p* = 0.1044, Figure 3.7C). The level of evidence to reject $H_0$ with groups compared to Spain R also decreased, down to no evidence with Spain L ($p = 0.1074$), weak evidence with Wales ($p = 0.0512$), and evidence with Oman ($p = 0.0457$).

## 3.9 DISCUSSION

Our analysis suggests the presence of a dominant movement pattern (morphotype 1) in *C. semiplicata*, with little difference incurred by a change in geographic location or temporary resting (i.e. rusophyciform depressions). This can be interpreted to suggest a dominant stereotyped behaviour was employed by all *C. semiplicata* tracemakers in the late Cambrian to early Ordovician. We hypothesized that the "different" specimens (i.e. those not belonging to morphotype 1) may likewise document variations of this behaviour which were also similar across groups. To further investigate this question, we examined the 40 "different" specimens in our specimen-vs-specimen two-sample *t*-test (Supplementary Information, Figure 3.13). Within this matrix, we observed a plaid or checkerboard patterning to the *p*-values, with alternating regions with no evidence to disprove $H_0$ ($p > 0.1$) and regions with weak to strong evidence to disprove H$_0$ ($p < 0.1$). Working from the regions of no evidence to disprove H$_0$ (i.e. black squares in Supplementary Information, Figure 3.13), we were able to further cluster the "different" specimens into two groups. When the *p*-value matrix is organized according to these groups (Figure 3.10), two overarching trends become clear: (1) that specimens predominantly show no evidence to disprove H$_0$ ($p > 0.1$) when compared with specimens of the same group and (2) that specimens predominately show weak or stronger evidence to disprove H$_0$ ($p < 0.1$) when compared with specimens of a different group.

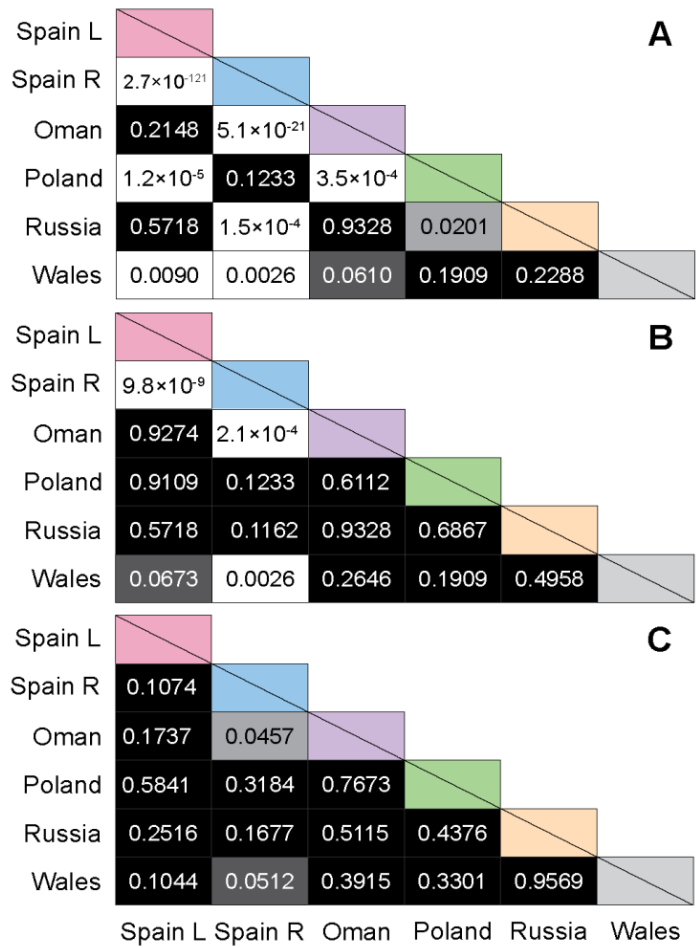

Figure 3.7. Results of group-vs-group two-sample t-tests performed with: A, grouped turning angle data (seg. mult = 0.5), not adjusted for preferential turning direction. B: grouped turning angle data (seg. mult = 0.5), adjusted for preferential turning direction. C: grouped turning angle data (seg. mult = 0.5) with outliers removed, adjusted for preferential turning direction. Shade of boxes reflect p-value ranges: black = no evidence ($p > 0.1$), dark grey = weak evidence ($0.1 > p > 0.05$), light grey = evidence ($0.05 > p > 0.01$), white = strong evidence ($p < 0.01$).

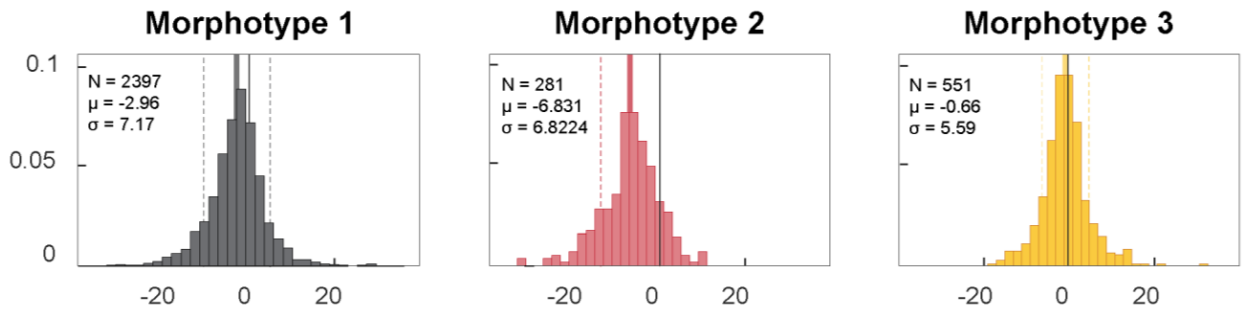

Figure 3.8. PDFs of direction-adjusted turning angle data for every specimen of morphotype 1, 2, and 3.

This suggests the presence of two more distinct morphotypes of C. semiplicata: Morphotype 2 and morphotype 3. Morphotypes 1, 2, and 3 potentially reveal distinct behavioural variants within C. semiplicata (Fig. 8). They would likely reflect small changes to the dominant stereotyped behaviour of the tracemaker, perhaps reflecting behavioural flexibility to changing external conditions (e.g. food availability), internal drives (e.g. change from primarily feeding to primarily locomotion), or motion capacity (e.g. impact of sediment consistency). To further examine the potential causes for these changes in behaviour, we looked at the presence of each morphotype on the Spanish slab (Fig. 9). Here, specimens of specific morphotypes tend to occur in similar vicinities. This is especially apparent in the specimens of morphotype 3 in Subgroup A and those of morphotype 2 in Subgroup D. In turn, specimens of morphotype 2 tend to be more curved, while those of morphotype 1 are more linear.

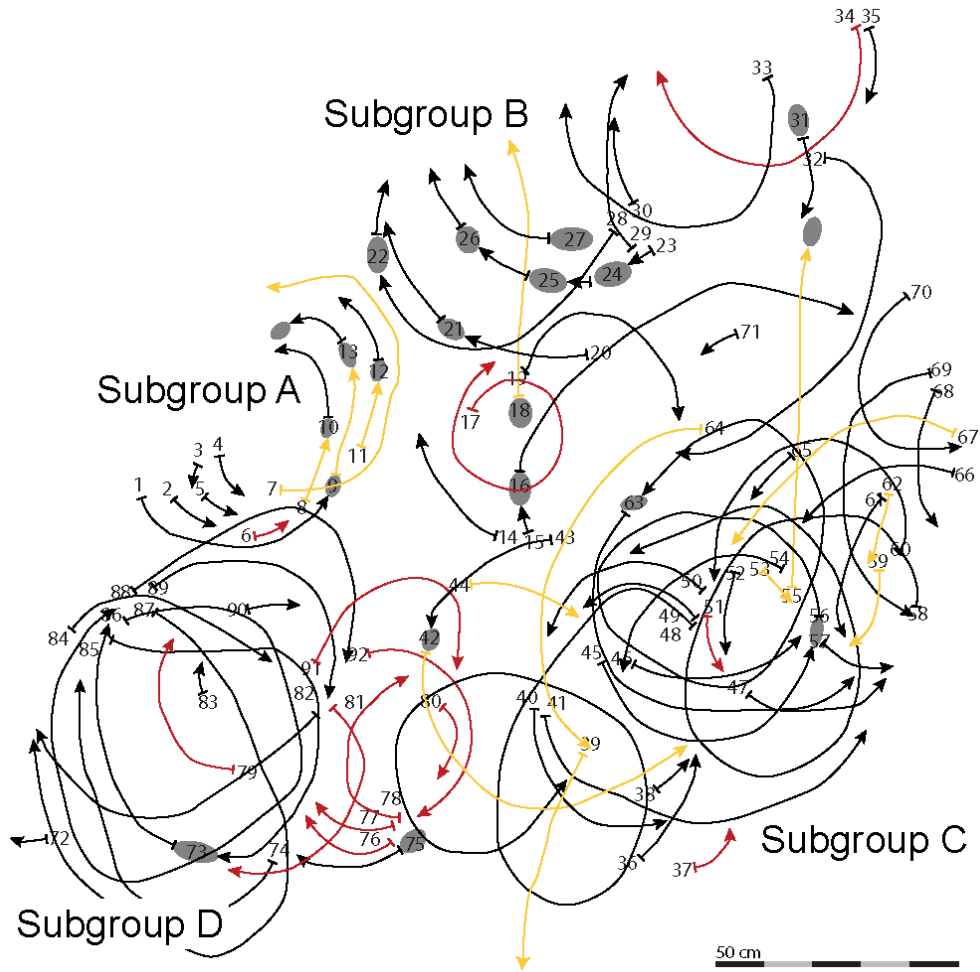

Figure 3.9. Schematic of *C. semiplicata* specimens from Spain with morphotypes 1 (black), 2 (red), and 3 (yellow) as well as rusophyciform depressions (dark grey) indicated.

Results from *t*-tests performed on locality-based groupings of turning data (e.g. subgroup-vs-subgroup tests of Spain A, B, C and D, or group-vs-group tests of Spain vs. Oman) revealed potential differences in their movement patterns. These differences subside, however, when only specimens demonstrating the dominant movement path (i.e. morphotype 1) are analyzed and grouped by locality. This suggests the original results of our subgroup-vs-subgroup and group-vs-group two-sample *t*-tests were affected by a disproportionate sampling of "different" movement patterns (i.e. morphotypes 2 and 3). When re-analyzed through the perspective of the three morphotypes revealed by the specimen-vs-specimen *t*-tests, this effect appears likely. Subgroups within Spain R (Spain B and D) contain a disproportionate number of trajectories belonging to morphotype 2, which likely accounted for the original observed differences.

3.10 CONCLUSIONS

This paper offers ichnologists a new addition to their investigative toolkit. With careful considerations of the constraints and limitations of fossil datasets, we developed a method to discretize fossil movement paths. Our method extracts spatial and inferred temporal data on fossil trajectories, allowing for the application of numerable statistical tests already widely applied on similar trajectories of extant organisms. It offers a way to statistically test hypotheses and quantitatively describe and compare fossil movement paths.

Our investigation of *Cruziana semiplicata* specimens from the late Cambrian to early Ordovician highlights this utility. The application of one statistical technique (two-sample t-tests) at varying scales of analysis (specimen-level, subgroup-level, and group-level) allowed for three previous assertions about *C. semiplicata* to be tested. Our results support Seilacher's inference that the *C. semiplicata* tracemaker was able to follow the same stereotyped behaviour after temporary resting and over considerable distances. Three morphotypes within *C. semiplicata* were revealed by our analysis, interpreted as one dominant stereotyped behaviour and two subordinate behaviours. We found little evidence that this dominant behaviour altered significantly across the groups, even though certain groups employed different subordinate behaviours more often. Notably, we found no obvious relationships between inferred paleoenvironment and prevalence of different subordinate behaviours. Higher-resolution analyses of associations between sedimentology, paleoenvironment, and *C. semiplicata* morphotypes may shed light on the relationship between external environment, motion capacity, and applied movement patterns. The phi distributions collected via our methodology offer an additional tool for describing trace fossil morphology and variability in ichnotaxonomic work. In turn, our results do not support separating *C. semiplicata* into two ichnosubspecies and instead reveal a mosaic of behavioural variants present in varying proportions across localities, perhaps reflecting changes in internal states or behavioural flexibility of the *C. semiplicata* tracemaker to differing external conditions.

These results highlight the utility of our method to morphologically examine trajectories within an ichnospecies. The same base methodology can also be used to examine how trajectories interact within a community, develop through geologic time, or alter across ecosystems. Are there statistical similarities across ichnospecies, ichnogenera, or ichnofamilies? What information can we extract about the navigation and motion capacities of the tracemaker? How does inferred paleoenvironment affect the movement paths within an ichnospecies? Is there a measurable effect on movement paths across times of ecological change, such as the Cambrian Explosion, Great Ordovician Biodiversification Event, Mesozoic Marine Revolution, or the Permian-Triassic Extinction Event? Are certain strategies employed more often in specific paleoenvironments (i.e. what is the extent of behavioural flexibility)? In addition to analyzing behavioural differences within *C. semiplicata*, our method offers a foundation to investigate these additional questions. The discretization of fossil movement paths presented here aligns fossil movement data with extant organism data, opening the wealth of statistical techniques applied in movement ecology research for use by ichnologists.

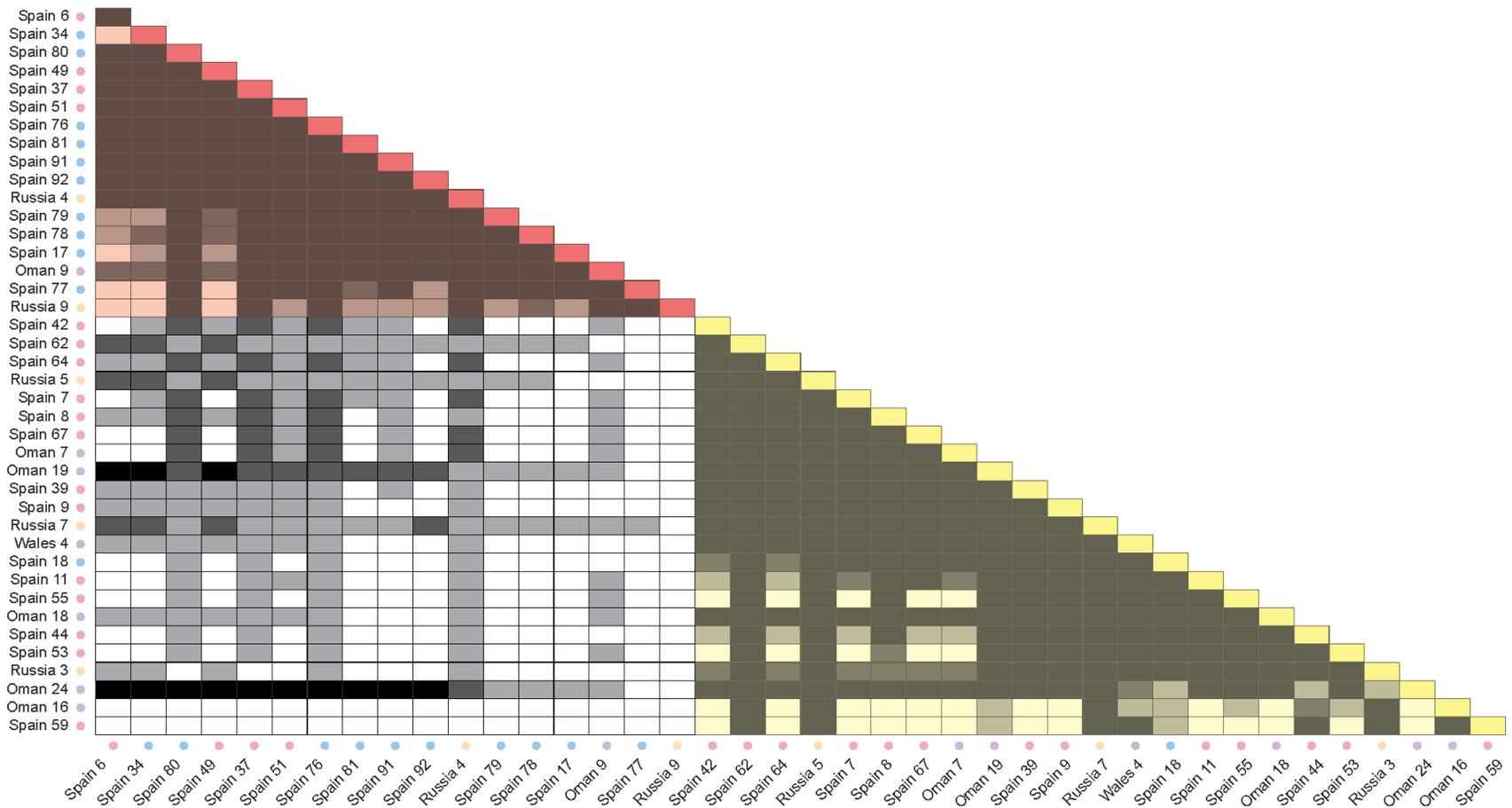

Figure 3.10. Specimen vs. specimen two-sample t-test results for all specimens that had p-values less than 0.1 with at least 25% of the total examined specimens, performed on direction-adjusted turning angle data (seg. mult = 0.5). Shade of boxes reflect p-value ranges: black = no evidence (p > 0.1), dark grey = weak evidence (0.1 > p > 0.05), light grey = evidence (0.05 > p > 0.01), while = strong evidence (p < 0.01). Yellow and red shading indicates specimens of different morphotypes. Small coloured circles indicate the associated group for each specimen: pink = Spain L, blue = Spain R, purple = Oman, orange = Russia, Grey = Wales.

## 3.11 ACKNOWLEDGEMENTS

B. A. L is supported by an NSERC PGS-D grant as well as an iMQRES scholarship from Macquarie University. M. G. M. and L. A. B. thank financial support provided by Natural Sciences and Engineering Research Council (NSERC) Discovery Grants 311727-15/20, and 311726–13 and 422931-20, respectively. M. G. M. also thanks funding by the George J. McLeod Enhancement Chair in Geology.

## 3.12 DECLARATION OF COMPETING INTERESTS

The authors declare none.

Fagan, W. F., Lewis, M. A., Auger-Méthé, M., Avgar, T., Benhamou, S., Breed, G., Ladage, L., Schlägel, U. E., Tang, W. W., Papastamatiou, Y. P., Forester, J., & Mueller, T. (2013). Spatial memory and animal movement. *Ecology Letters*, *16*(10), 1316–1329. https://doi.org/10.1111/ele.12165

Fan, R., Uchman, A., & Gong, Y. (2017). From morphology to behaviour: Quantitative morphological study of the trace fossil *Helminthorhaphe*. *Palaeogeography, Palaeoclimatoglogy, Palaeoecology, 485*, 946–955. https://doi.org/10.1016/j.palaeo.2017.08.013

Fitch, A. (1850). A historical, topographical and agricultural survey of the County of Washington. *Transactions of the New York Agricultural Society*, *9*, 753–944.

Fortey, R. A., & Seilacher, A. (1997). The trace fossil *Cruziana semiplicata* and the trilobite that made it. *Lethaia, 30*(2), 105–112. https://doi.org/10.1111/j.1502-3931.1997.tb00450.x

Hofmann, H. J. (1990). Computer simulation of trace fossils with random patters, and the use of goniograms. *Ichnos 1*(1), 15–22. https://doi.org/10.1080/10420949009386327

Hofmann, H. J., & Patel, I. M. (1989). Trace fossils from the type 'Etcheminian Series' (Lower Cambrian Ratcliffe Brook Formation), Saint John area, New Brunswick, Canada. *Geological Magazine*, *126*(2), 139–157. https://doi.org/10.1017/S0016756800006294

Jensen, S., Bogolepova, O. K., & Gubanov, A. P. (2011). *Cruziana semiplicata* from the Furongian (Late Cambrian) of Severnaya Zemlya Archipelago, Arctic Russia, with a review of the spatial and temporal distribution of this ichnospecies. *Geological Journal, 46*, 26–33. https://doi.org/10.1002/gj.1248

Jensen, Z. A. (2017). *Behavioural Paleoecology of Lower Cambrian Deposit Foragers : Reinterpreting Looping and Meandering Traces using Optimal Foraging Theory and Quantitative Analysis* (Publication No. 10273785) [Master's thesis, University of Nevada]. ProQuest Dissertations Publishing.

Jones, A. R. E. (1977). Search Behaviour : A Study of Three Caterpillar Species. *Behaviour, 60*(3–4), 237–259. https://www.jstor.org/stable/4533802

Joo, R., Picardi, S., Boone, M. E., Clay, T. A., Patrick, S. C., Romero-Romero, V. S., & Basille, M. (2022). Recent trends in movement ecology of animals and human mobility. Movement *Ecology, 10*(26), 1–20. https://doi.org/10.1186/s40462-022-00322-9
82

3.14 SUPPLEMENTARY INFORMATION

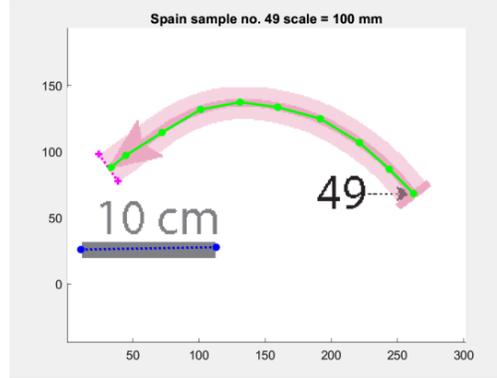

Figure 3.11. Example of the segmentation process of discretized fossil movement paths. Green line denotes the fossil movement path, with filled circles indicating equidistant points along the path. Pink + and blue points indicate points used to calculate the width of the fossil path and scale bar respectively.

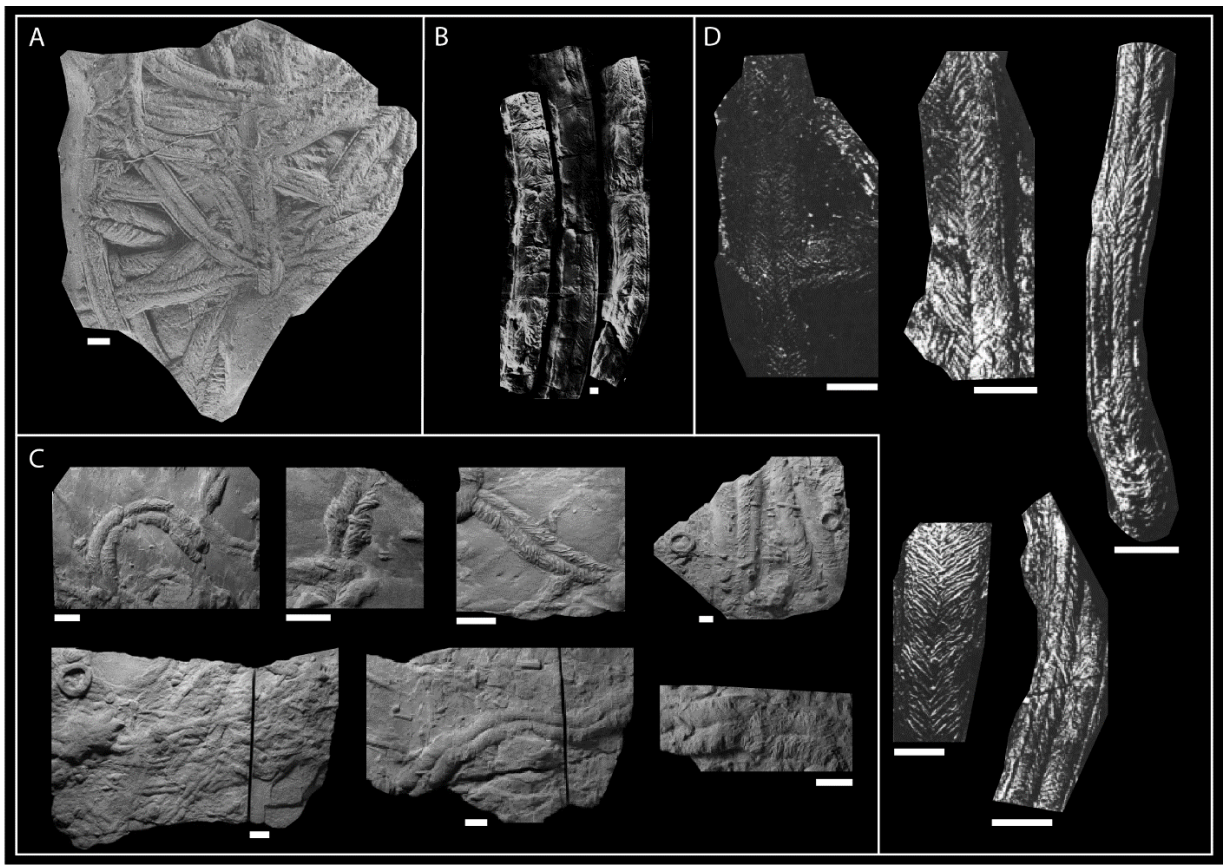

Figure 3.12. Specimens from (A) Oman (Fortey & Seilacher, 1997; Figure 1) (B) Poland (Radwański and Roniewicz, 1972; Figure 1) (C) Russia (Jensen et al., 2011: Figure 2 & Figure 3) and (D) Wales (Crimes, 1968; 1970, Plate 5) used in the analysis. All scale bars are 1 cm.

Supplementary Figure 3.13. Specimen vs. specimen two-sample *t*-test results for all *C. semiplicata* specimens, performed on direction-adjusted turning angle data (seg. mutl. = 0.5). Yellow lines delineate different thresholds.

Table 3.2. Subgroup summary data. SI= straightness index.

| Subgroup | Avg. trail width (mm) | Total trail length (mm) | Mean SI | No. specimens | Group |
|---|---|---|---|---|---|
| Spain A | 16.5012 | $2.6211 \times 10^3$ | 0.9174 | 13 | Spain L |
| Spain C | 24.7105 | $1.9426 \times 10^4$ | 0.8605 | 36 | Spain L |
| Spain B | 32.2089 | $8.1844 \times 10^3$ | 0.8280 | 22 | Spain R |
| Spain D | 37.6909 | $1.0427 \times 10^4$ | 0.7346 | 21 | Spain R |
| Oman | 9.7156 | $1.0800 \times 10^4$ | 0.977 | 24 | Oman |
| Poland | 41.8986 | $1.0790 \times 10^3$ | 0.9823 | 3 | Poland |
| Russia | 10.4538 | 767.0834 | 0.9304 | 12 | Russia |
| Wales | 12.0706 | 60.3531 | 0.9882 | 5 | Wales |

Table 3.3. All specimens included in the *C. semiplicata* analysis.

| Batch ID | Group ID | S ID | Width | Scale (mm) | Scale (pixels) | Scale ratio | Length (mm) | SI |
|---|---|---|---|---|---|---|---|---|
| Spain | SpainA | 1 | 16.58 | 100 | 100.53 | 0.99 | 459.24 | 0.83 |
| Spain | SpainA | 2 | 16.79 | 100 | 100.14 | 1.00 | 106.03 | 0.98 |
| Spain | SpainA | 3 | 15.84 | 100 | 100.75 | 0.99 | 46.66 | 1.00 |
| Spain | SpainA | 4 | 15.74 | 100 | 99.88 | 1.00 | 98.49 | 0.95 |
| Spain | SpainA | 5 | 16.21 | 100 | 100.20 | 1.00 | 73.66 | 0.99 |
| Spain | SpainA | 6 | 16.66 | 100 | 99.90 | 1.00 | 76.07 | 0.98 |
| Spain | SpainA | 7 | 16.25 | 100 | 100.27 | 1.00 | 801.02 | 0.52 |
| Spain | SpainA | 8 | 17.42 | 100 | 99.05 | 1.01 | 132.93 | 1.00 |
| Spain | SpainA | 9 | 16.52 | 100 | 99.94 | 1.00 | 222.34 | 0.98 |
| Spain | SpainA | 10 | 15.35 | 100 | 99.14 | 1.01 | 206.81 | 0.84 |
| Spain | SpainA | 11 | 17.49 | 100 | 100.59 | 0.99 | 133.05 | 1.00 |
| Spain | SpainA | 12 | 16.91 | 100 | 99.98 | 1.00 | 155.34 | 0.96 |
| Spain | SpainA | 13 | 16.74 | 100 | 101.47 | 0.99 | 123.15 | 0.89 |
| Spain | SpainB | 14 | 30.83 | 100 | 99.69 | 1.00 | 270.42 | 0.96 |
| Spain | SpainB | 15 | 32.69 | 100 | 100.07 | 1.00 | 51.85 | 1.00 |
| Spain | SpainB | 16 | 32.71 | 100 | 94.69 | 1.06 | 896.72 | 0.88 |
| Spain | SpainB | 17 | 31.42 | 100 | 100.14 | 1.00 | 816.97 | 0.13 |
| Spain | SpainB | 18 | 33.79 | 100 | 101.88 | 0.98 | 520.51 | 0.98 |
| Spain | SpainB | 19 | 35.25 | 100 | 99.62 | 1.00 | 513.08 | 0.65 |
| Spain | SpainB | 20 | 32.70 | 100 | 102.12 | 0.98 | 246.16 | 0.99 |
| Spain | SpainB | 21 | 33.20 | 100 | 100.30 | 1.00 | 223.69 | 0.99 |
| Spain | SpainB | 22 | 33.20 | 100 | 100.40 | 1.00 | 108.73 | 0.99 |
| Spain | SpainB | 23 | 32.94 | 100 | 100.57 | 0.99 | 53.72 | 1.00 |
| Spain | SpainB | 24 | 31.37 | 100 | 101.50 | 0.99 | 44.55 | 1.00 |
| Spain | SpainB | 25 | 32.61 | 100 | 101.02 | 0.99 | 111.25 | 0.99 |
| Spain | SpainB | 26 | 32.53 | 100 | 101.03 | 0.99 | 129.22 | 0.99 |
| Spain | SpainB | 27 | 31.58 | 100 | 100.80 | 0.99 | 245.15 | 0.92 |
| Spain | SpainB | 28 | 30.15 | 100 | 101.47 | 0.99 | 643.64 | 0.74 |
| Spain | SpainB | 29 | 28.49 | 100 | 99.11 | 1.01 | 375.52 | 0.94 |
| Spain | SpainB | 30 | 31.38 | 100 | 100.59 | 0.99 | 179.27 | 0.98 |
| Spain | SpainB | 31 | 33.26 | 100 | 99.84 | 1.00 | 171.98 | 0.96 |
| Spain | SpainB | 32 | 30.85 | 100 | 98.93 | 1.01 | 903.83 | 0.76 |
| Spain | SpainB | 33 | 34.26 | 100 | 100.04 | 1.00 | 811.08 | 0.52 |
| Spain | SpainB | 34 | 32.17 | 100 | 99.81 | 1.00 | 706.18 | 0.59 |
| Spain | SpainB | 35 | 31.20 | 100 | 101.79 | 0.98 | 152.95 | 0.97 |
| Spain | SpainC | 36 | 23.68 | 100 | 101.79 | 0.98 | 239.20 | 0.97 |
| Spain | SpainC | 37 | 24.12 | 100 | 99.71 | 1.00 | 113.11 | 0.92 |
| Spain | SpainC | 38 | 23.31 | 100 | 99.26 | 1.01 | 91.21 | 1.00 |
| Spain | SpainC | 39 | 24.86 | 100 | 99.63 | 1.00 | 456.45 | 0.98 |
| Spain | SpainC | 40 | 23.29 | 100 | 100.64 | 0.99 | 423.91 | 0.83 |
| Spain | SpainC | 41 | 26.00 | 100 | 101.81 | 0.98 | 809.50 | 0.80 |
| Spain | SpainC | 42 | 25.28 | 100 | 100.24 | 1.00 | 774.85 | 0.73 |
| Spain | SpainC | 43 | 25.41 | 100 | 100.59 | 0.99 | 325.02 | 0.94 |
| Spain | SpainC | 44 | 27.60 | 100 | 101.44 | 0.99 | 229.75 | 0.98 |
| Spain | SpainC | 45 | 25.69 | 100 | 99.73 | 1.00 | 603.87 | 0.72 |
| Spain | SpainC | 46 | 25.16 | 100 | 100.90 | 0.99 | 362.46 | 0.93 |
| Spain | SpainC | 47 | 26.58 | 100 | 101.03 | 0.99 | 312.17 | 0.90 |
| Spain | SpainC | 48 | 24.44 | 100 | 100.57 | 0.99 | 2563.81 | 0.16 |
| Spain | SpainC | 49 | 24.55 | 100 | 102.31 | 0.98 | 257.83 | 0.87 |
| Spain | SpainC | 50 | 26.20 | 100 | 101.22 | 0.99 | 403.12 | 0.81 |
| Spain | SpainC | 51 | 24.55 | 100 | 100.18 | 1.00 | 117.09 | 0.96 |
| Spain | SpainC | 52 | 24.99 | 100 | 99.68 | 1.00 | 165.05 | 0.99 |
| Spain | SpainC | 53 | 24.43 | 100 | 99.64 | 1.00 | 88.10 | 1.00 |
| Spain | SpainC | 54 | 23.44 | 100 | 101.88 | 0.98 | 451.74 | 0.85 |
| Spain | SpainC | 55 | 25.64 | 100 | 97.83 | 1.02 | 713.17 | 1.00 |
| Spain | SpainC | 56 | 25.89 | 100 | 101.49 | 0.99 | 1977.37 | 0.22 |
| Spain | SpainC | 57 | 23.20 | 100 | 101.25 | 0.99 | 157.55 | 0.95 |
| Spain | SpainC | 58 | 22.53 | 100 | 101.80 | 0.98 | 2092.97 | 0.28 |
| Spain | SpainC | 59 | 24.06 | 100 | 99.12 | 1.01 | 179.61 | 0.94 |
| Spain | SpainC | 60 | 25.66 | 100 | 99.90 | 1.00 | 739.95 | 0.52 |
| Spain | SpainC | 61 | 25.21 | 100 | 100.10 | 1.00 | 311.33 | 0.95 |
| Spain | SpainC | 62 | 23.87 | 100 | 100.27 | 1.00 | 152.20 | 0.99 |
| Spain | SpainC | 63 | 25.62 | 100 | 100.69 | 0.99 | 888.40 | 0.65 |
| Spain | SpainC | 64 | 23.35 | 100 | 100.79 | 0.99 | 851.25 | 0.81 |
| Spain | SpainC | 65 | 25.49 | 100 | 101.25 | 0.99 | 109.66 | 0.99 |
| Spain | SpainC | 66 | 24.26 | 100 | 99.56 | 1.00 | 353.54 | 0.93 |
| Spain | SpainC | 67 | 24.94 | 100 | 101.29 | 0.99 | 562.92 | 0.90 |
| Spain | SpainC | 68 | 25.96 | 100 | 100.48 | 1.00 | 291.34 | 0.94 |
| Spain | SpainC | 69 | 22.97 | 100 | 99.74 | 1.00 | 661.56 | 0.77 |
| Spain | SpainC | 70 | 22.34 | 100 | 101.08 | 0.99 | 504.11 | 0.65 |
| Spain | SpainC | 71 | 25.00 | 100 | 100.47 | 1.00 | 83.90 | 0.99 |
| Spain | SpainD | 72 | 36.89 | 100 | 102.35 | 0.98 | 71.46 | 0.99 |
| Spain | SpainD | 73 | 36.96 | 100 | 101.76 | 0.98 | 506.79 | 0.92 |
| Spain | SpainD | 74 | 37.75 | 100 | 100.99 | 0.99 | 799.67 | 0.68 |
| Spain | SpainD | 75 | 36.60 | 100 | 100.60 | 0.99 | 218.13 | 0.96 |
| Spain | SpainD | 76 | 39.05 | 100 | 99.66 | 1.00 | 214.24 | 0.86 |
| Spain | SpainD | 77 | 38.89 | 100 | 99.81 | 1.00 | 182.04 | 0.91 |
| Spain | SpainD | 78 | 38.69 | 100 | 99.68 | 1.00 | 418.64 | 0.71 |
| Spain | SpainD | 79 | 37.43 | 100 | 101.69 | 0.98 | 353.52 | 0.81 |
| Spain | SpainD | 80 | 36.16 | 100 | 99.71 | 1.00 | 166.44 | 0.88 |
| Spain | SpainD | 81 | 38.99 | 100 | 101.47 | 0.99 | 534.68 | 0.74 |
| Spain | SpainD | 82 | 40.24 | 100 | 101.86 | 0.98 | 712.71 | 0.79 |
| Spain | SpainD | 83 | 39.39 | 100 | 101.03 | 0.99 | 64.49 | 0.97 |
| Spain | SpainD | 84 | 36.72 | 100 | 100.23 | 1.00 | 473.52 | 0.89 |
| Spain | SpainD | 85 | 39.09 | 100 | 101.21 | 0.99 | 912.93 | 0.54 |
| Spain | SpainD | 86 | 33.54 | 100 | 101.36 | 0.99 | 1699.24 | 0.02 |
| Spain | SpainD | 87 | 36.84 | 100 | 100.13 | 1.00 | 1515.60 | 0.24 |
| Spain | SpainD | 88 | 38.51 | 100 | 102.49 | 0.98 | 737.79 | 0.61 |
| Spain | SpainD | 89 | 36.67 | 100 | 98.95 | 1.01 | 602.42 | 0.72 |
| Spain | SpainD | 90 | 36.22 | 100 | 100.60 | 0.99 | 107.62 | 0.97 |
| Spain | SpainD | 91 | 39.53 | 100 | 101.19 | 0.99 | 524.55 | 0.55 |
| Spain | SpainD | 92 | 37.35 | 100 | 99.94 | 1.00 | 523.14 | 0.69 |
| Oman | Oman | 1 | 8.75 | 10 | 68.97 | 0.14 | 44.98 | 0.98 |
| Oman | Oman | 2 | 9.22 | 10 | 69.69 | 0.14 | 133.64 | 0.90 |
| Oman | Oman | 3 | 8.92 | 10 | 66.43 | 0.15 | 103.95 | 0.97 |
| Oman | Oman | 4 | 9.62 | 10 | 69.06 | 0.14 | 51.86 | 0.97 |
| Oman | Oman | 5 | 11.36 | 10 | 67.73 | 0.15 | 98.06 | 0.92 |
| Oman | Oman | 6 | 9.82 | 10 | 66.82 | 0.15 | 186.99 | 0.71 |
| Oman | Oman | 7 | 9.39 | 10 | 70.36 | 0.14 | 112.81 | 0.98 |
| Oman | Oman | 8 | 11.52 | 10 | 66.58 | 0.15 | 30.96 | 0.98 |
| Oman | Oman | 9 | 9.17 | 10 | 66.35 | 0.15 | 28.25 | 0.96 |
| Oman | Oman | 10 | 9.37 | 10 | 68.36 | 0.15 | 84.65 | 0.90 |
| Oman | Oman | 11 | 8.96 | 10 | 69.44 | 0.14 | 17.73 | 0.99 |
| Oman | Oman | 12 | 9.12 | 10 | 67.79 | 0.15 | 36.94 | 0.99 |
| Oman | Oman | 13 | 8.87 | 10 | 68.74 | 0.15 | 13.33 | 0.98 |
| Oman | Oman | 14 | 9.07 | 10 | 68.35 | 0.15 | 14.77 | 1.00 |
| Oman | Oman | 15 | 13.49 | 10 | 69.46 | 0.14 | 86.44 | 0.99 |
| Oman | Oman | 16 | 10.02 | 10 | 66.82 | 0.15 | 48.79 | 0.94 |
| Oman | Oman | 17 | 9.03 | 10 | 69.03 | 0.14 | 99.60 | 0.96 |
| Oman | Oman | 18 | 9.32 | 10 | 66.84 | 0.15 | 49.45 | 1.00 |
| Oman | Oman | 19 | 8.48 | 10 | 69.70 | 0.14 | 41.24 | 0.99 |
| Oman | Oman | 20 | 9.21 | 10 | 68.23 | 0.15 | 34.47 | 0.99 |
| Oman | Oman | 21 | 10.84 | 10 | 66.82 | 0.15 | 148.66 | 0.99 |
| Oman | Oman | 22 | 9.09 | 10 | 69.02 | 0.14 | 9.42 | 1.00 |
| Oman | Oman | 23 | 11.96 | 10 | 70.92 | 0.14 | 49.42 | 0.99 |
| Oman | Oman | 24 | 10.38 | 10 | 68.08 | 0.15 | 21.65 | 1.00 |
| Poland | Poland | 1 | 41.99 | 10 | 14.61 | 0.68 | 341.67 | 0.99 |
| Poland | Poland | 2 | 44.83 | 10 | 14.05 | 0.71 | 412.95 | 0.98 |
| Poland | Poland | 3 | 38.88 | 10 | 16.01 | 0.62 | 331.30 | 0.98 |
| Russia | Russia | 1 | 16.35 | 5 | 58.59 | 0.17 | 78.34 | 0.99 |
| Russia | Russia | 2 | 15.90 | 20 | 87.46 | 0.11 | 48.53 | 0.99 |
| Russia | Russia | 3 | 15.31 | 10 | 90.98 | 0.11 | 52.78 | 1.00 |
| Russia | Russia | 4 | 3.56 | 20 | 103.27 | 0.10 | 14.08 | 0.90 |
| Russia | Russia | 5 | 3.68 | 20 | 104.10 | 0.10 | 34.69 | 0.98 |
| Russia | Russia | 6 | 3.24 | 20 | 103.56 | 0.10 | 41.73 | 0.94 |
| Russia | Russia | 7 | 3.17 | 20 | 102.12 | 0.10 | 9.62 | 0.99 |
| Russia | Russia | 8 | 3.38 | 10 | 102.17 | 0.10 | 25.86 | 0.96 |
| Russia | Russia | 9 | 11.01 | 10 | 88.50 | 0.11 | 70.01 | 0.58 |
| Russia | Russia | 10 | 21.34 | 5 | 67.05 | 0.15 | 70.14 | 0.98 |
| Russia | Russia | 11 | 14.30 | 10 | 53.01 | 0.19 | 115.36 | 0.93 |
| Russia | Russia | 12 | 14.22 | 10 | 96.56 | 0.10 | 56.80 | 0.92 |
| Wales | Wales | 1 | 11.91 | 6 | 82.27 | 0.07 | 51.69 | 0.98 |
| Wales | Wales | 2 | 12.88 | 6 | 83.28 | 0.07 | 63.52 | 0.99 |
| Wales | Wales | 3 | 10.06 | 6 | 80.72 | 0.07 | 77.92 | 0.97 |
| Wales | Wales | 4 | 10.90 | 6 | 81.60 | 0.07 | 51.96 | 1.00 |
| Wales | Wales | 5 | 14.60 | 6 | 83.57 | 0.07 | 44.58 | 0.99 |